\documentstyle[floats,multicol,aps,epsf,prl]{revtex}

\begin{document}
\draft
\twocolumn[\hsize\textwidth\columnwidth\hsize\csname @twocolumnfalse\endcsname
\title{
Two lifetimes in the thermopower of the cuprate metals
}
\author{J. A. Clayhold$^*$ and Z. H. Zhang}
\address{Texas Center for Superconductivity, University of Houston,
Houston, TX 77204-5932}
\author{A. J. Schofield}
\address{Theory of Condensed Matter, Cavendish Laboratory, Madingley Road, 
Cambridge CB3 0HE, United Kingdom}
\date{\today}
\maketitle
\begin{abstract}
We study the temperature-dependence of thermoelectric transport in
proton-irradiated thin films of Tl$_2$Ba$_2$CaCu$_2$O$_{8+\delta}$.
All the anomalous transport conductivities are affected by the
irradiation but maintain their non-Fermi liquid forms. We note however
that the anomalous temperature dependence of the thermopower scales
precisely with the temperature dependent Hall coefficient. This
provides strong evidence that the two relaxation rates observed to
control magnetotransport are also responsible for the anomalous
temperature dependence of the thermopower.
\end{abstract}

\vskip 0.2 truein
\vskip2pc
]

Understanding the normal state transport properties of the cuprate
superconductors remains one of the major challenges posed by these
materials. A key insight has been the notion of two distinct
relaxation rates controlling transport processes, which has proved to
be an important unifying concept in quantifying the anomalies in
magnetotransport~\cite{pwa}.  It addresses the temperature-- and
impurity--dependences of the Hall coefficient~\cite{chien} as well as
the violation of Kohler's rule~\cite{harris}. In this Letter we
present evidence that the anomalous temperature dependence of the
thermopower is simply related to that of the Hall coefficient. The
implication is that the same two relaxation rates which govern
magnetotransport are also responsible for the unusual behavior of the
thermopower.

The experimental evidence for two qualitatively different relaxation
rates (inverse lifetimes) in the normal state is compelling.  The
resistivity, $\rho_{xx}$, increases linearly with the temperature at
optimal doping which optical measurements~\cite{forro} indicate to be
due to a scattering rate, $\Gamma_{tr}$, proportional to the
temperature, $T$,
\begin{equation}
\Gamma_{tr} \sim \eta T \quad (\eta \sim 2) \; .
\label{gtr}
\end{equation}
Doping with scattering impurities like zinc leads to an additional
residual term~\cite{cooper1} in $\Gamma_{tr}$ in keeping with
Matthiessen's rule for a scattering rate.  Changes in the carrier
concentration away from optimal doping cause the temperature
dependence of $\Gamma_{tr}$ to evolve in a complex manner, as does the
temperature dependence of the Hall coefficient, $\rm R_H$. By
contrast, their ratio $(\rho_{xx}/{\rm R_H})$---the inverse Hall
angle---maintains a remarkably robust
form~\cite{chien,carrington1,carrington2} and also shows Matthiessen's
rule behavior as the impurity concentration, $n_i$, is changed.
Anderson interpreted this as a new scattering rate, $\Gamma_H$, which
governs the decay of Hall currents~\cite{pwa}
\begin{equation}
\Gamma_H = {T^2 \over W} + n_i b \; .
\label{gH}
\end{equation}
These two rates must appear multiplicatively in the Hall conductivity
$\sigma_{xy} \sim \omega_c/\Gamma_{tr} \Gamma_H$.  Optical
measurements~\cite{kaplan} seem to confirm this form for
$\sigma_{xy}$. This second rate, $\Gamma_H$, also controls the
magneto-resistance\cite{harris} accounting for the failure of $\Delta
\rho_{xx}(B)/\rho_{xx}(0)$ to scale with $B/\rho_{xx}$. To date
however no one has looked for empirical evidence of these two rates
outside magnetotransport. Here we study the thermopower which is the
other in-plane transport property which shows systematic
deviations~\cite{kaiser,obertelli,cooper2} from the usual Drude metal
form.

Electric currents can be driven by both
electric fields and thermal gradients,
\begin{equation}
j^e_x = \sigma_{xx} E_x + \beta_{xx} \nabla_x T \; .
\end{equation}
The thermopower, being the electric field per unit temperature
gradient under conditions of electrical isolation, is given by the
ratio of the thermoelectric and electric conductivities,
$S=-\beta_{xx}/\sigma_{xx}$.  For the diffusion thermopower of a Drude
metal we know that
\begin{equation}
\sigma_{xx} = {\omega_p^2 \over \Gamma} \; , \quad \quad
\beta_{xx} = -{e T {\cal L}_0 \omega_p^2 \over \epsilon_F \Gamma} \; ,
\end{equation}
where $\omega_p=ne^2/m$ is the plasma frequency and ${\cal L}_0 =
\pi^2 k_B^2/3e^2$ is the Lorentz number. Thus the scattering rates
normally cancel in the diffusion thermopower and $S/T \sim e{\cal
L}_0/\epsilon_F$ is independent of temperature. In the cuprates this
relation is obeyed only in the heavily overdoped regime.  Here we
investigate whether the anomalous temperature dependence of the
thermopower could be understood within the two relaxation rate
scenario. We know that $\Gamma_{tr}$ is, by definition, the scattering
rate controlling $\sigma_{xx}$, but the scattering rate which should
appear in $\beta_{xx}$ is not, {\it a priori}, known. However we only
require the temperature dependence of the dimensionless ratio of that
scattering rate with $\Gamma_{tr}$. The only such parameter one can
make out of the two rates is $\Gamma_{tr}/\Gamma_H$ so on dimensional
grounds we must have
\begin{equation}
\label{scaling}
{S/T} = f[\Gamma_{tr}/\Gamma_H] = f[{\rm R_H}(T)/\omega_c] \;,
\end{equation}
{\it i.e.} that $S/T$ is a function of the Hall number.

Of course the above analysis makes many assumptions, in particular
that the scattering rates are not strongly energy dependent and the
same conduction processes enter both electrothermal and
galvanomagnetic conduction. An explicit realization of this scaling
relation is given by the phenomenological transport equation of
Coleman, Schofield and Tsvelik~\cite{coleman}. From symmetry
considerations they argue that to discriminate between Hall and
electric currents the scattering mechanism must be sensitive to the
charge conjugation symmetry of the quasiparticles.  Using their
expressions for the transport conductivities~\cite{details} we can
obtain an explicit form for the scaling function, $f$,
\begin{equation}
\label{CSTform}
{S(T) \over T} = {e {\cal L}_0 \over 2 \epsilon_F} \left[ 1 +
{\omega_p^2 \over \omega_c} {\rm R_H(T)} \right] \; .
\end{equation}
It is expected that the coefficients in this linear relation should
depend strongly on the carrier concentration.

Experimentally testing Eqs.~\ref{scaling} and \ref{CSTform} is a
delicate affair because of the extreme sensitivity of the thermopower
in the cuprates to very small changes of the carrier density.  Small
changes in $p$, the number of holes/Cu at the level of
${\Delta}p$~=~0.001 are significant enough to change the entire
thermopower curve~\cite{obertelli}. In principle, Eqs.~\ref{scaling}
and~\ref{CSTform} could be studied using a series of Zn-- or Ni--doped
samples to vary the scattering lifetimes.  It would, however, be
nearly impossible to ascertain the oxygen-content of the various
chemically-doped samples with sufficient accuracy to keep fixed the
coefficients in Eq.~\ref{CSTform}.  Measurements made on a series of
chemically-doped samples would also be susceptible to uncertainty from
any lack of reproducibility in the geometries of the electrical and
thermal contacts.

Thus we see that in order to experimentally test Eq.~\ref{CSTform}, or
the more general scaling form of Eq.~\ref{scaling}, we need to fix the
carrier doping and change only the scattering rates. This can be
achieved by introducing point defects using proton irradiation of thin
film samples.  We used a commercially-available\cite{STI} thin film of
Tl$_2$Ba$_2$CaCu$_2$O$_{8+\delta}$. Tl$_2$Ba$_2$CaCu$_2$O$_{8+\delta}$
is particularly well-suited for this study because it is tetragonal,
with no Cu-O chain contribution to the thermoelectric power.  We
measured the resistivity, thermopower, Hall effect, and Nernst
effect\cite{TBP} for temperatures ranging from T$_c$ to 400~K in the
as-grown sample.  The sample was subsequently irradiated with 400 keV
protons to a dose of $10^{16}$ p/cm$^2$, creating a uniform density of
point-defects throughout the 1 {$\mu$}m thick sample.  The sample was
then re-measured with the same contacts, thermocouples and heater
intact from the first measurement.  It was possible using the
irradiation in this way to change the scattering lifetimes without
affecting either the sample stoichiometry or the contact geometries.

\begin{figure}[tb]
\epsfxsize=3.5in
\centerline{\epsfbox{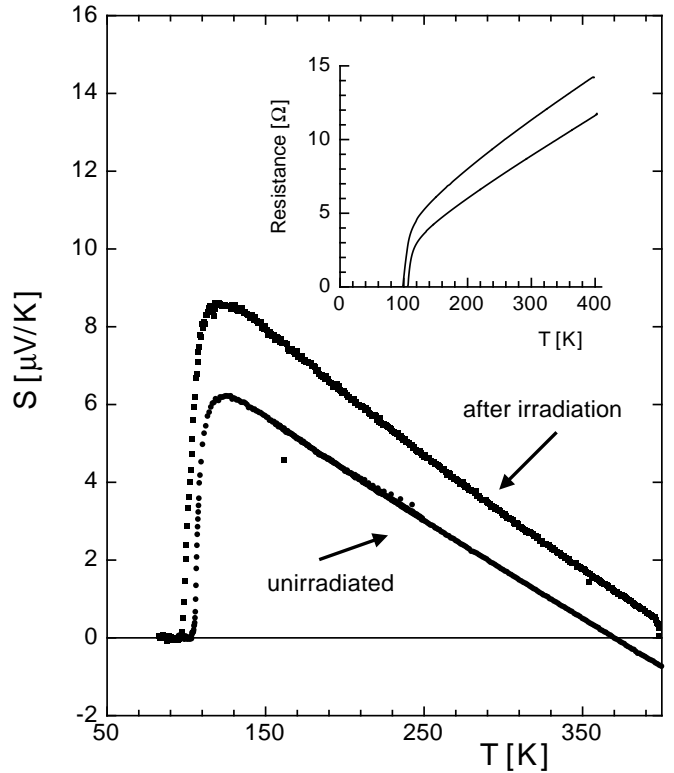}}
\protect\caption{
The thermopower before and after proton irradiation. As discussed in
the text, the proton bombardment not only increased the residual
resistivity as expected, but also had a dramatic effect on the
thermopower but without changing the carrier doping.
\label{tepresis}}
\end{figure}

Results are shown in Fig.~\ref{tepresis}.  The inset shows the
resistance before and after irradiation with the upper curve showing
the increased resistance after the proton bombardment.  Contrary to
expectation for a Fermi liquid, the irradiation had a marked effect on
the thermopower, increasing it by nearly 2 $\mu$V/K. As discussed
below, the upward shift of the thermopower curve appears to be an
intrinsic lifetime effect, not attributable for example, to a change
of carrier density from the irradiation.

\begin{figure}[tb]
\epsfxsize=3.5in
\centerline{\epsfbox{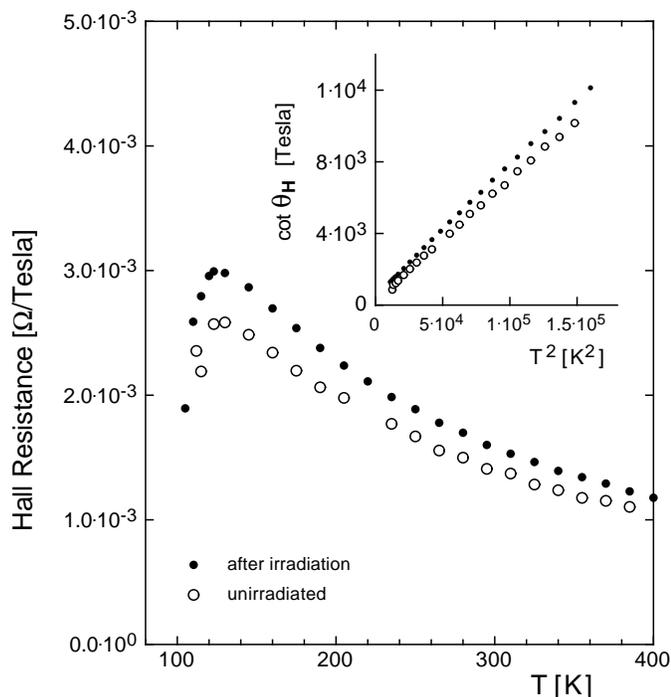}}
\protect\caption{
The Hall resistivity and Hall angle before and after proton
irradiation. The Hall angle showed the usual quadratic temperature
dependence but was not very dramatically changed by proton
irradiation.
\label{Hallfig}}
\end{figure}

Results for the Hall effect, showing the usual $1/T$ behavior, before
and after irradiation are shown in Fig.~\ref{Hallfig}.  The Hall angle
(inset) yielded a straight line when plotted against the square of the
temperature both before and after irradiation.  It is noteworthy that
the intercept of the Hall angle curves, the residual Hall scattering,
was only negligibly affected by the increased density of point
defects, unlike the longitudinal residual resistance.  The nature of
the scatterer (point defect, Zn substitution, Ni substitution)
apparently affects the two scattering channels differently.

\begin{figure}[tb]
\epsfxsize=3.5in
\centerline{\epsfbox{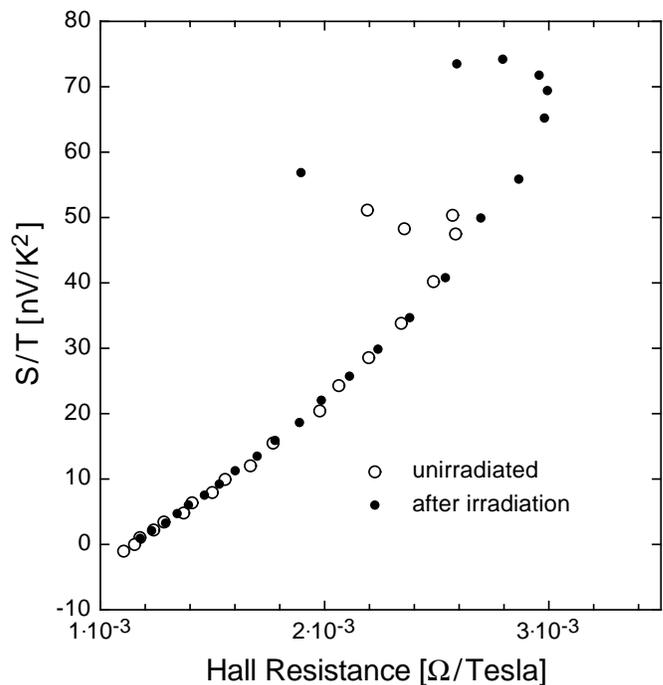}}
\protect\caption{
Combining the results of Fig.~\ref{tepresis} and Fig.~\ref{Hallfig}
using the scaling relation of Eq.~\ref{scaling}, we see that the
curves before and after irradiation coincide. This indicates that the
two relaxation rates $\Gamma_{tr}$ and $\Gamma_H$ determine the
anomalous temperature dependence of both the Hall coefficient {\em
and} the thermopower. Deviations occur near the superconducting
fluctuation regime.
\label{scalefig}}
\end{figure}

When, according to Eqs.~\ref{scaling} and \ref{CSTform}, we compare
the thermopower and the Hall effect by plotting S/T as a function of
$R_H$, the differences in the measurements before and after the proton
bombardment all but vanish, as seen in Fig.~\ref{scalefig}.  The two
curves lie exactly on top of each other, diverging only in the region
of superconducting fluctuations.  This is the natural consequence of
Eqs.~\ref{scaling} and \ref{CSTform}, in which all dependence on
lifetimes is subsumed into the appearance of $R_H(T)$.  The
coefficients of the linear relation, Eq.~\ref{CSTform}, for example,
can only depend on bandwidths, carrier concentrations, etc., but {\em
not} the relaxation rates.  Since the band structure is presumably not
altered significantly by the small increase of point defects, these
coefficients should be the same before and after the irradiation.
This is what is observed.  The slight upward curvature indicates that
the general form, Eq.~\ref{scaling} may be more applicable at lower
temperatures, nearer to T$_c$.

For consistency with the thermopower systematics observed by
Obertelli, Cooper and Tallon\cite{obertelli}, the slope of the curve
in Fig.~\ref{scalefig} must be extremely sensitive to changes of
carrier concentration.  If the proton bombardment had depleted or
added carriers, the agreement of the two curves in Fig.~\ref{scalefig}
would have been highly unlikely.  This result lends confidence that
the experimental data were sensitive only to lifetime effects, as
originally intended.

We now discuss the implications for such a scaling relation in the
normal state. The obvious implication is that the two scattering rates
are intrinsic to the normal state and, in particular, it is not
necessary to have a magnetic field in order to observe the
consequences of the Hall relaxation rate. Two microscopic mechanisms
which rely on the magnetic field in this way would appear to be
disfavored by this result: the skew scattering mechanism of Kotliar
{\it et al}~\cite{kotliar} and the gauge model picture of Lee and
Lee~\cite{lee}. In the former, the presence of the magnetic field
invokes a new, singular, scattering mechanism.  In the latter picture
the spin--fermions and charge--bosons interact strongly via a fictitious
transverse gauge field which effectively screens any applied magnetic
field. It is only the real electrons which can couple to the magnetic
field and feel the Lorentz force so the cyclotron frequency acquires a
temperature dependent factor which modifies the Hall angle.

An addition to showing that the two rates are intrinsic to the normal
state in the absence of a magnetic field, this experiment has
implications for the mechanism that determines which scattering rate
is active in a given transport experiment.  It has been customary to
associate $\Gamma_{tr}$ with longitudinal currents: namely those which
change the energy of the quasiparticle distribution (like the current
driven by an electric field). The Hall relaxation rate has been
associated with transverse currents which involve a rotation of the
quasiparticle distribution without a change of energy (like the Hall
component of the current). An example of such a parameterization is
achieved by replacing $\omega_c \tau$ by $\omega_c \tau_H$ in the
usual Drude conductivities~\cite{harris}.  In this experiment we are
seeing evidence of the Hall relaxation rate in the thermopower which
is a longitudinal current. Thus the distinction between longitudinal
and transverse currents is not sufficient to determine which
relaxation rate should be seen experimentally.

One common feature shared by both the the Hall current and the
diffusion thermo-electric current is their dependence on the inverse
effective mass tensor. The Hall current is determined by the curvature
of the Fermi surface---the transverse effective mass. The diffusion
thermopower requires hot electrons to move faster than cold holes and
so is sensitive to the longitudinal effective mass. This suggests that
$\tau_H$ might well be associated with any term involving the inverse
effective mass, not just $\omega_c$.

Finally we comment on theoretical pictures which predict such a
relation. Anderson originally introduced the two relaxation times
based on a picture of spin--charge separation in the normal state: the
two lifetimes reflecting the time taken by an electron to decay into a
spinon and holon ($\Gamma_{tr}^{-1}$) and the lifetime of the spinons
($\Gamma_H^{-1}$).  He has also suggested~\cite{pwabook} that these
two lifetimes might show up in the thermopower based on studies of
spin-charge separation in one dimension~\cite{stafford}. Cooper and
Carrington have suggested a connection between the thermopower and the
Hall coefficient based on a picture of anisotropic
scattering~\cite{cooper3}. Our motivation for considering a scaling
relation came from the two-lifetimes phenomenology of Coleman,
Schofield and Tsvelik~\cite{coleman}. In their picture the reason
$\Gamma_H$ enters both the Hall conductivity and the thermoelectric
conductivity is indeed related to their dependence on the inverse
effective mass tensor.  (Currents proportional to the inverse
effective mass transform under charge conjugation with the opposite
sign to currents depending only on Fermi velocity.)  The scaling form
we find in Fig.~\ref{scalefig} is very close to the linear relation
that their phenomenology would predict (Eq.~\ref{CSTform}). However,
there is one discrepancy, namely the relative sign between the two
scattering rates. Their phenomenology predicts that $\beta_{xx} \sim
\Gamma_{tr}^{-1} + \Gamma_{H}^{-1}$ {\it i.e.}
like two fluids with thermoelectric conductivities adding.
Experimentally we see that the two conductivities subtract (indicated
by the positive gradient but negative intercept in
Fig.~\ref{scalefig}). Thus it is as if we have two fluids which
carry opposite signs of thermo-electric current. It is not clear that
this discrepancy can be cured by the inclusion of a realistic Fermi
surface in Ref.~\onlinecite{coleman}.

In summary, we have have studied the effect of proton irradiation on
the in-plane normal state transport properties of $Tl2212$ thin films
for a wide range of temperatures. We have demonstrated a link between
the anomalous temperature dependence of the thermopower and the Hall
coefficient. This suggests that the two relaxation rates used to
quantify the anomalies in galvomagnetic transport are also responsible
for those seen in thermoelectric transport. We have discussed the
implications of this result for a number of microscopic pictures of
the normal state.

We have benefitted from informative discussions with Q. Si,
P. W. Anderson and J. L. Tallon and one of us (AJS) would acknowledge
the hospitality of Rice University where this work was begun.

\noindent
{\footnotesize $^*$Present Address: Department of Physics and Astronomy,
Clemson University, Clemson, SC 29634-1911.}

\end{document}